\documentclass[prd,preprint,12pt]{revtex4}
\pdfoutput=1
\usepackage{hyperref}
\usepackage{graphicx}
\usepackage{amsmath,amssymb,amsfonts,amsthm,latexsym,stmaryrd}
\usepackage{marginnote}
\usepackage{color}
\usepackage{soul}

\newcommand{\FK}{K}

\newcommand{\cC}{\sigma}
\newcommand{\C}{A}

\newcommand{\cR}{\kappa}  
\newcommand{\Lm}{L_m} 

\begin{document}   
\title{Axial perturbations of neutron stars with shift symmetric conformal coupling}
\author{Hamza Boumaza}\email{boumaza14@yahoo.com, boumaza.hamza@univ-jijel.dz}    
\affiliation{\small Laboratory of Theoretical physics (LPTH) and Department of Physics, faculty of Exact and Comuter Science, Mohamed Seddik Ben Yayia university, Bp 98 Ouled Aissa, Jijel 18000, Algeria}
 \begin{abstract}
{\hskip 2em}In this present work, the axial quasi-normal modes of neutron stars, with a shift symmetric conformal coupling, are studied for different realistic equations of state.  First, we derive the background equations in static and spherically symmetric spacetime and then we solve them numerically by taking into account the continuity and regularity conditions. Second, we extend our calculus to the perturbed level where we derive the equations of motion as well as the ghost and Laplacian instability. We find that the proposed model is free from these instabilities everywhere. We find also that the variation of the quasi-normal modes is affected by the conformal coupling where the deviation from general relativity is observed. Finally, with this motivation, we fit our numerical result with universal relations for axial quasi-normal modes using five type of realistic equations of state. The universality of  the scaled frequency and damping time in terms of the compactness in this model is confirmed in this model.
 \end{abstract} 

\maketitle

\section{Introduction}
{\hskip 2em} In the last years, the structure of the neutron stars (NSs) has became one of the most interesting puzzle that excites astrophysicists \citep{Berti:2015itd,Koivisto:2012za,Kase:2020qvz,Babichev:2016jom,Saito:2015fza}. In fact, probing the nature of the matter inside the core of a neutron star, in our laboratory, is very difficult, since its central density is larger than $10^{15}g/cm^3$ which can not be realized on earth.  Also, when we are close to the surface of a NS, the gravity is very strong which is also impossible to realize it in our laboratories. Therefore, astronomical observations on compact objects are essential to have a better understanding on the behaviours of matter at high density and the gravity in strong regime.   

{\hskip 2em}Studying the mass and the radius of this compact object can give us important information on the equation of state (EOS) at these extreme conditions. However, even the mass can be measured with high precision, the determination of radius is quit difficult with a direct astrophysical observation. Nevertheless, the detection of gravitational waves (GWs), from the collision of a binary neutron star or from the collision of a binary Black Holes by  the LIGO-VIRGO Collaboration \cite{Abbott_2017,abbott2018gw170817,armengol2017neutron,LIGOScientific:2017vwq}, opens a new era of astronomy to understand the physical features of compact objects. The determination of the speed propagation of GWs, from the events GW170817 \cite{LIGOScientific:2017vwq} followed by a short gamma ray burst \cite{Goldstein_2017}, has loured out many alternative theory of gravity, where Degenerate-Higher-Order-Scalar-Tensor theories \cite{Langlois:2018dxi} have been reduced considerably \cite{Ezquiaga:2017ekz}.  Moreover, simulating the tidal love number with GWs observations from a BNSs collision can also reduced a considerable number of the EOSs \cite{Flanagan:2007ix}. Although, the advance that our detector have made, we still need more precision, more signals and more theoretical studies to comprehend the remnant object (Black Holes, neutron stars....etc).

{\hskip 2em}The Quasi-Normal Modes (QNMs), which are  small, complex and nonradial oscillations defined by the general relativistic equations, sound to be  promising quantities for the future GWs detection. They were first studied by Thorne and Campolattaro \cite{thorne1967} and then by Lindblom and Detweiler  \cite{Lindblom:1983ps,Detweiler:1985zz}, where they take into account the fluid and gravitational oscillations. In the last Refs, the authors neglected the imaginary part  against the real part in order  to derive the QNMs of a purely outgoing waves. These modes, called $p$ and $f$ modes, can also be found in classical gravity. Then, Kokkotas and Schutz discovered a new families of normal modes in which the imaginary part is not negligible by using WKB-approximation, and they called it $w$ modes \cite{kokkotas1992w}. However, this approximation is not viable for large imaginary part, therefore the authors of Ref.\cite{leins1993nonradial} used the continued fraction method to prove the existence of a new branch of strongly damped normal modes, called  $w_{II}$ modes. More details about the classification  of QNMs can be found in Ref.\cite{Kokkotas:1999bd}.

{\hskip 2em}The quasi-normal modes have been studied in the context of general relativity, but one can investigate the implications of alternative theory of gravity on the QNMs. This issue has been studied for a black hole in Ref.\cite{Langlois:2021aji}, in which the authors have considered a particular case of Horndeski theory \cite{horndeski1974second}. The axial and the polar QNMs have been derived and compared to general relativity (GR). The axial modes of neutron stars  were also studied in $R^2$-gravity \cite{Blazquez-Salcedo:2018qyy}, where the results differ from GR by varying  the parameter $a$. They have derived the universal relations in $R^2$-gravity, shown in Refs.\cite{Blazquez-Salcedo:2018tyn,Blazquez-Salcedo:2013jka,Blazquez-Salcedo:2012hdg}. In Ref.\cite{AltahaMotahar:2019ekm}, the axial QNMs of neutron stars in scalar-tensor theories with massive scalar field including a self-interacting term in the potential have been investigated for various realistic equations of state. Like in Ref. \cite{Blazquez-Salcedo:2018qyy}, different universal relations was determined to show the effects of the scalar field on the QNMs of neutron stars. The coupling of the scalar field with the matter through a conformal or disformal transformation \cite{Bekenstein:1992pj} have also been investigated in other literature to describe the characteristic of a relativistic stars \cite{Ikeda:2021skk,Boumaza:2021hzr,Minamitsuji:2016hkk}.

{\hskip 2em}In this article, we consider our previous model of the Ref.\cite{Boumaza:2021hzr} where the matter is coupled to the scalar field and the metric through a shift symmetric conformal transformation.   Where  we have found that the scalar field is vanished outside the star which allow us to derive the polar and axial TLNs of the NS. This interesting feature will be used to calculate the axial QNMs and thus enables us to identify the difference between this model and GR. We restrict our study only on the axial perturbations since they have no similar modes in Newtonian gravity and are a pure manifestation of the tensorial character of GR. 

{\hskip 2em}This article is organized as follow: In the next section, we review   the general equations of motion and  the general form of the energy-momentum tensors in the Jordan frame and in the Einstien from. We will also review  the Tolman-Oppenheimer-Volkoff (TOV) equations and we will show our numerical results of these equations. In section \ref{Sec2}, we present the axial perturbations and we develop the perturbed equations for a particular case. We will also present the asymptotic behaviours of the axial metric as well as the conditions of ghost and Laplacian instability. In section \ref{Sec3}, we show our numerical method that we have used to extract the axial QNMs and then we discuss the numerical results by using the universal relations. Finally, we end the paper with a conclusion.

\section{Background equations:}\label{Sec1}
{\hskip 2em} In this section, we will review the model studied in \cite{Boumaza:2021hzr} and we will focus on the background equations. To have simple mathematical calculations, we derive the background equations of motion directly from the total action of the model by using Euler-Lagrange equations. Finally, we show the numerical solutions and proprieties of the neutron star in which the matter is coupled to metric trough a conformal transformation.
\subsection{The model:}\label{Sec11}
 {\hskip 2em}Let's consider a  model described by a \textit{geometrical} (or gravitation) metric $g_{\mu\nu}$ and a \textit{physical} metric $\overline{g}_{\mu\nu}$ (the matter in the universe follows the geodesics of \textit{physical} metric), where the metrics $g_{\mu\nu}$ and  $\overline{g}_{\mu\nu}$ are, respectively, the Einstien and Jordan frame metrics and they are linked by  the Conformal transformation 
\begin{eqnarray}
\overline{g}_{\mu\nu}=\C(X)\; g_{\mu\nu},\label{conformal transformation}
\end{eqnarray}
where $\C$ (called Conformal factor) is an arbitrary functions of  the kinetic term $X=\varphi^{;\mu}\varphi_{;\mu}$, where ";" notation represents the covariant derivative $_{;\mu}\equiv\nabla_{\mu}$ and $\varphi$ is the scalar field. Note that the signature $(-,+,+,+)$ is conserved and  the causality is respected only if  this function is positive \cite{Bekenstein:1992pj}. Note that it is difficult to find  the inverse transformation because of the term $X$ which contains the metric of Einstein frame.  We will limit our study to shift-symmetric  case, e.i. the action is invariant under the transformation $\varphi\rightarrow\varphi + const$, as well as we will use the special function proposed in \cite{Boumaza:2021hzr}
\begin{eqnarray}
\C(X)=1+ \cC X^{\frac{1}{2}}.\label{function}
\end{eqnarray}

where $\cC$ is a real constant. Now, let's suppose a theory of gravity described by the following action  
\begin{eqnarray}
S=S_{vac}+S_{m}=\int d^4 x\;\left\lbrace\sqrt{-g}\;L_{vac}+\sqrt{-\overline{g}}\; \Lm(\overline{g}_{\mu\nu},\psi)\right\rbrace,\quad \text{with}\quad L_{vac}=\frac{\cR}{2}R+\FK (X),\label{S}
\end{eqnarray}
where  $R$ is the Ricci scalar in Einstien frame, with $\cR = (8 \pi G)/c^4$ ($c$ is the speed of light and $G$ is the gravitational coupling in Einstein frame), describes the gravitational sector. $L_m$ correspond to  the Lagrangian density of the  matter field, $\psi$, in the meanwhile the dynamic of the scalar field is described by  the arbitrary function $\FK$. Our model is not only characterized by shift symmetric Lagrangian  but it also  verify the constraint placed by detection of gravitational waves due to a collision of BNSs (GW170817) \cite{LIGOScientific:2017vwq}, followed by a short gamma ray burst \cite{Goldstein_2017}, where the propagation speed of GWs is equivalent to the speed of light \cite{Ezquiaga:2017ekz}. The model can be extend by adding Horndeski  or DHOST terms in $L_{vac}$ and by generalizing the transformation (\ref{conformal transformation}) to disformal transformation. But this feature will not considered in this paper. 
\subsection{Energy-momentum tensor:}\label{Sec12}
We wish to study the matter inside of neutron star which can expressed by the contravariant energy-momentum tensor in Jordan frame as follow
\begin{eqnarray}
\overline{T}^{\mu\nu(m)}=\frac{2}{\sqrt{-\overline{g}}}\frac{\delta(\sqrt{-\overline{g}}L_m)}{\delta\overline{g}_{\mu\nu}}=(\overline{\rho}+\overline{P})\overline{u}^\mu\overline{u}^\nu +\overline{g}^{\mu\nu}\overline{P},\label{Tmunu}
\end{eqnarray}
where $\overline{P}$ and $\overline{\rho}$ are the pressure and the energy density of the fluid in Jordane frame. The four vector velocity of the fluid, denoted by $\overline{u}^\mu$,  is constrained by $\overline{u}^\mu\overline{u}_\mu=-1$. The bar means that the quantity is expressed in Jordan frame while  we omit the bar for  Einstien frame. Hence, we define the tensor energy-momentum, in this frame, by 
\begin{eqnarray}
T^{\mu\nu(m)}=\frac{2}{\sqrt{-g}}\frac{\delta(\sqrt{-\overline{g}}L_m)}{\delta g_{\mu\nu}}=(\rho + P)u^\mu u^\nu +g^{\mu\nu}P,\label{TmunuE}
\end{eqnarray}
where $\rho$, $P$ and $u^\mu$ are the energy density, the pressure and the four velocity of the matter present in the universe. Using the definitions (\ref{conformal transformation}) and (\ref{Tmunu}), the relation between the tensors $T^{\mu\nu(m)}$ and $\overline{T}^{\mu\nu(m)}$ can be written as
\begin{eqnarray}
T^{\mu\nu(m)}=\sqrt{\frac{\overline{g}}{g}}\frac{\delta \overline{g}_{\alpha\beta}}{\delta g_{\mu\nu}}\overline{T}^{\alpha\beta(m)}=\C^3 \overline{T}^{\mu\nu(m)}.\label{TmunuE2}
\end{eqnarray}

Provided that the matter action is invariant under the coordinate transformation $x^\mu\rightarrow x^\mu+ \xi^\mu$, the energy-momentum tensor is covariantly conserved in Jordane frame \cite{Zumalacarregui:2013pma}. And thus the conservation equation of the fluid in this frame reads
\begin{eqnarray}
\overline{\nabla}_{\mu}\overline{T}^{\mu\nu(m)}=0,\label{dtmunu}
\end{eqnarray}
where  the covariant derivative is  constructed using $\overline{g}_{\mu\nu}$ and its derivatives. 
However, in Einstien frame, the energy-momentum tensor is not conserved ($\nabla_{\mu}T^{\mu\nu(m)}=I^{\nu}$) due to the coupling that appeared in the transformation (\ref{conformal transformation}). The term $I^{\nu}$ can calculated by using the full expression of the covariant derivative $\overline{\nabla}_{\mu}$ of the equation (\ref{TmunuE}) and  the relations in Appendix. A of Ref.\cite{Zumalacarregui:2013pma}. if we do so, it follows that
\begin{eqnarray}
I^{\nu}=\frac{\C_X}{2\C}\left(T^{(m)}\left(\partial^\nu X-\varphi_{;\alpha}\varphi^{;\alpha\nu}\right)+6\varphi^{;\alpha}\varphi_{;\alpha\beta}T^{\beta\nu(m)}\right),\label{dTmunuE}
\end{eqnarray}
where $T^{(m)}=g_{\mu\nu}T^{\mu\nu(m)}$ is the trace of energy-momentum tensor. However, because of Bianchi identities, the covariant derivative of the total energy-momentum tensor of the model is covariantly conserved in Einstein frame.
\subsection{TOV-equations and numerical solutions:}\label{Sec13}
In order to study the gravitational behaviours inside and outside a neutron start, we first consider a static and spherically symmetric spacetime in Einstein frame, endowed with the metric
\begin{eqnarray}
ds_{background}^2 = -f(r)^2\, dt^2+ h(r)^2\, dr^2+ q(r)^2 r^2\,(d\theta^2 +\sin^2 \theta \, d\phi^2),\label{dsE}
\end{eqnarray}
where $f(r)$, $h(r)$ and $q(r)$ are functions of radial coordinate $r$. Second, in the two frame, we impose that the scalar field $\varphi(r)$, the pressure and the energy density are radial functions as well as the spatial components of four velocity vector are vanished. The transformations between
the two frames is commented in \cite{Boumaza:2021hzr} and it can be summarized as 
\begin{eqnarray}
& f(r)\rightarrow\frac{\overline{f}(r)}{\sqrt{\C(X)}},\qquad h(r)\rightarrow\frac{\overline{h}(r)}{\sqrt{\C(X)}},\qquad q(r)\rightarrow\frac{\overline{q}(r)}{\sqrt{\C(X)}},\nonumber\\
& \rho(r)\rightarrow\frac{\overline{\rho}(r)}{\C(X)},\qquad P(r)\rightarrow\frac{\overline{P}(r)}{\C(X)},\qquad  u^t\rightarrow\frac{\overline{u}^t}{\sqrt{\C(X)}}.
\end{eqnarray}
Hence, one can find that the vector $\overline{u}^t=1/(f(r)\sqrt{\C})$. To avoid complex equations, it is convenient to express  the gravitational part and the matter part, respectively, in Einstein frame and Jordan frame. Inserting the metric (\ref{dsE}) in $L_{vac}$ and after integrating by part the action $\int d^4 x \sqrt{-\overline{g}}L_{vac}$, it follows that
\begin{eqnarray}
S_{vac}=\int dr r^2 f h \sin (\theta )  \left\lbrace \cR \left( \frac{q'^2}{h^2 q^2}+\frac{2 f'q'}{f h^2 q}-\frac{2 q'}{r h^2 q}+\frac{2 h'}{r h^3}+\frac{1}{r^2q^2}-\frac{1}{r^2h^2}\right)+\FK\left[\left(\frac{\varphi'}{h}\right)^2\right] \right\rbrace ,\nonumber \\
\label{Lvac backgound}
\end{eqnarray}
where the prime "$'$" denotes the derivative with respect to $r$. Varying the action $S$ with respect to $f$, $h$ and $q$ then taking  the limits $q\rightarrow 1$, $q'\rightarrow 0$ and $q''\rightarrow 0$, we obtain the following equations
\begin{eqnarray}
\frac{h'}{h}&= & \frac{1-h^2}{2 r}-\frac{r h^2 \left(\FK-\C^2 \overline{\rho}\right)}{2 \cR},\label{eh}\\
\frac{f'}{f}&= & \frac{r  \left(\varphi'(r)^2 \left(\C \C_{X}(\overline{\rho}(r)-3  \overline{P})-2 \FK_{X}\right)+h^2 \left(\C^2 \overline{P}+\FK\right)\right)}{2 \cR}+\frac{ \left(h^2-1\right)}{2 r},\label{ef}\\
\frac{f''}{f}&= &\frac{h^2 \left(\C^2 \overline{P}+\FK\right)}{\cR }+\frac{h'}{r h}+\frac{f'}{f}\left(\frac{h'}{h}-\frac{1}{r}\right).\label{eff}
\end{eqnarray}
And the equation of  matter conservation reads
\begin{eqnarray}
\frac{\overline{P}'}{\overline{P}+\overline{\rho}}&=& \frac{\C_{X}}{\C h^2}\left(\frac{h' }{h} \varphi '^2-\varphi ' \varphi ''\right)-\frac{f'}{f}.\label{eP}
\end{eqnarray}
One can show that the last four equations are not independent because Eq. (\ref{eff}) can be obtained from the combination of the Eqs. (\ref{eh}), (\ref{ef}) and (\ref{eP}).
The scalar field equation is also obtained by varying the total action with respect to $\varphi$ which is calculated by
\begin{eqnarray}
\frac{d}{dr}\left[f h r^2 J\right]&=&0,\label{ephi2}
\end{eqnarray}
where
\begin{eqnarray}
J=\frac{\varphi'}{h^2}\left(\C \C_{X}(\overline{\rho}-3\overline{P})+2 \FK_{X}\right).\label{eJ}
\end{eqnarray}
Note that the typo that appears in the expression of $J^r$ of \cite{Boumaza:2021hzr} is corrected.  If we consider $J=0$ is a solution to th Eq. (\ref{ephi2}), one can deduce that $\varphi '=0$ or $\C \C_{X}(\overline{\rho}-3\overline{P})+2 \FK_{X}=0$. The first root corresponds to GR and the second one allows us to write the quantity $X$ as a function of the trace of the tensor energy-momentum $\overline{T}^{(m)}=\overline{\rho}-3\overline{P}$. This feature is important to have a vanishing Noether current at the center of star and thus $J$ is regular everywhere. Considering  the case (\ref{function}) and taking into account $\FK=X$, yields
\begin{eqnarray}
\varphi' =\frac{\cC h (3 \overline{P}-\overline{\rho} )}{4-\cC^2 (3 \overline{P} -\overline{\rho})}.\label{varphi'}
\end{eqnarray}
We see that the scalar radial derivative of scalar field is vanished when $\overline{P}=0$ and $\overline{\rho}$, e.i. we have $\varphi'=0$,  and thus $\varphi= const$, outside the star.  Therefore, due to this feature the Einstien frame and Jordan frame become identical in the vacuum. In addition, the equations of the metric field, outside the neutron stars, are equivalent to those in GR which are given by
\begin{eqnarray}
\overline{f}=f=\frac{f_{\infty}}{\overline{h}}=\frac{f_{\infty}}{h}=f_{\infty}\left(1-\frac{2 M}{r}\right)^{\frac{1}{2}},\label{extsol}
\end{eqnarray}
where $f_{\infty}$ is a real constant of integration and $M$ is defined as the mass of star. The constant $f_{\infty}$ can be imposed to be equal to $1$ or one can redefine the time coordinate as $t_p =t/f_{\infty}$.
\subsection{Numerical analysis:}\label{Sec13}
In order to  solve the equations of motion, we will focus on the particular case showed in the last subsection. To achieve the numerical integration, it is convenient to reduce our equations by replacing the equation (\ref{varphi'}) in  (\ref{eh}) and (\ref{eP}) which gives
\begin{eqnarray}
\frac{h'}{h}&=&\frac{1-h^2}{2 r}+\frac{h^2 r \left(16 \rho -\cC^2 (\rho -3 P)^2\right)}{2 \cR \left(\cC^2 (\rho -3 P)+4\right)^2},\label{h'/h}\\
\frac{\overline{P}'}{\overline{P}+\overline{\rho}}&=&-\frac{2c_m^2   \left(4-\cC^2 (\overline{\rho} -3  \overline{P})\right)}{ \left(8c_m^2+\cC^2\left( \overline{\rho} -5 c_m^2 \overline{\rho}+\overline{P}+3 c_m^2 \overline{P}\right)\right)}\frac{f'}{f},
\end{eqnarray}
where $c_m^2\equiv\partial\overline{P}/\partial\overline{\rho}$ is the speed sound of the matter inside the neutron star. In addition,  the function $f$ can be obtained by integrating the following equation
\begin{eqnarray}
\frac{f'}{f}=\frac{h^2-1}{2 r}+\frac{h^2 r\left(16 \overline{P}-3 \cC^2 (\overline{\rho} -3 \overline{P})^2\right)}{2 \cR  \left(4+\cC^2 (\overline{\rho}-3 \overline{P})\right)^2}.
\end{eqnarray}
Moreover, we need the equation of state $\overline{P}(\overline{\rho})$, where we will use the realistic  EOSs simulated from binary neutron star coalescence, defined by
\begin{eqnarray}
\zeta &=& \frac{\text{a}_1+\text{a}_2 \xi +\text{a}_3 \xi ^3}{(\text{a}_4 \xi +1)}d\left(\text{a}_5 (\xi -\text{a}_6)\right)+(\text{a}_7+\text{a}_8 \xi)d\left(\text{a}_9 (\text{a}_{10}-\xi )\right)+(\text{a}_{11}+\text{a}_{12} \xi)d\left(\text{a}_{13} (\text{a}_{14}-\xi )\right)\nonumber\\
& & +(\text{a}_{15}+\text{a}_{16} \xi )d\left(\text{a}_{17} (\text{a}_{18}-\xi )\right)
 +\frac{\text{a}_{19}}{\text{a}^2_{20} (\text{a}_{21}-\xi )^2+1}+\frac{\text{a}_{22}}{\text{a}^2_{23} (\text{a}_{24}-\xi )^2+1},\label{EOS}
\end{eqnarray} 
with
\begin{eqnarray}
d(x)=\frac{1}{e^x+1}.
\end{eqnarray}
where $\xi=log_{10}(\overline{\rho}/g\;cm^{-3})$ and $\zeta=log_{10}(\overline{P}/g\;cm^{-3})$. The value of the coefficients $a_i$ can be found in  Refs.\cite{Haensel:2004nu,Potekhin:2013qqa}. \\

\begin{figure}[h]
\centering
\includegraphics[scale=0.82]{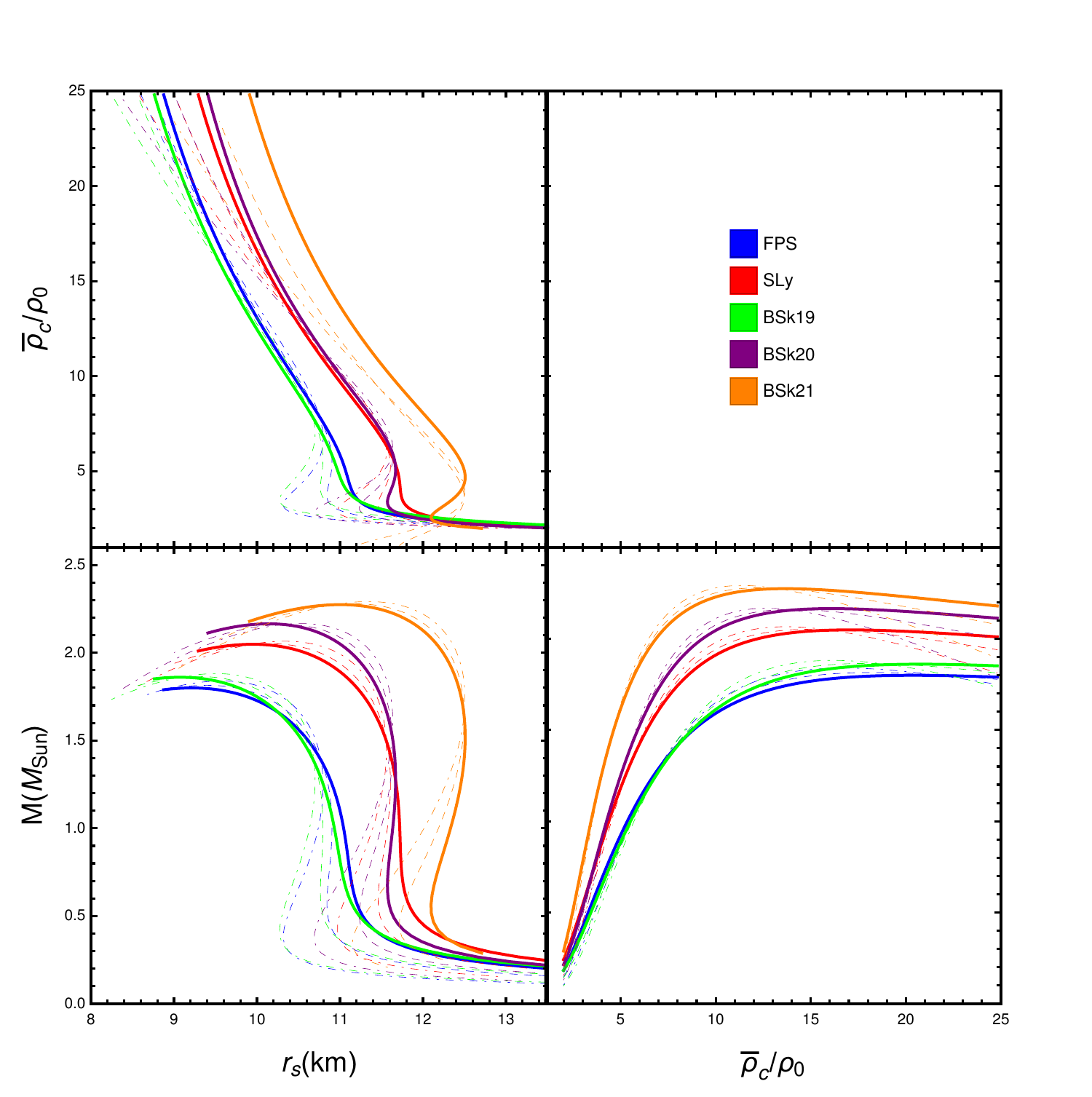} 
\caption{\small The mass-Radius-central energy density relations of NS   using the EOSs: FPS, SLy, BSk19, BSk20 and BSk21 for: $\Tilde{\cC}=0$ (Solide lines),  $\Tilde{\cC}=0.03$ (Dashed  lines),  $\Tilde{\cC}=0.045$ (DotDashed  lines).  }
\label{MRs}
\end{figure}

The numerical integration of the equations above, from the center to the surface, is performed by  using the dimensionless variable $s=\ln(r/r_0)$, where $r_0=c/\sqrt{G \rho_0}=89.664$ and $\rho_0=1.6749\times 10^{14}\; g.cm^{-3}$, and by redefining $\cC$ and $\varphi$ as: $\cC\rightarrow \tilde{\cC}/(\cR\rho_0^2)$ and $\varphi\rightarrow\sqrt{\cR}\varphi$. At the center of the star we impose the regularity conditions in order to have stable solutions when $r\rightarrow 0$. Then, the constants of integration in (\ref{extsol}) are determined by matching the exterior and the interior solutions, for more details see \cite{Boumaza:2021hzr}. In addition, from the numerical solutions, one can derive the radius of the star using the equation: $\overline{P}(r_s)=0$ for each  value of the central energy density $\overline{\rho}_c$. In other words, the radius and the mass depend not only   the equation of state but they are also affected by on the value of $\Tilde{\cC}$ and  $\overline{\rho}_c$. As an example, we show the impact of constant $\Tilde{\cC}$ on the relations mass, radius and central energy density for different kind of EOSs, in Fig.\ref{MRs}.  

\section{Axial perturbations:}\label{Sec2}
{\hskip 2em}This section is devoted to study the axial perturbations around a static and spherical symmetric background. In fact, we will derive and analyze the equations of motion inside and outside the star. In addition, the conditions of Ghost and Laplacian instability of the theory given by the action (\ref{S}) are also derived. The polar perturbations are not considered in this paper, but we leave this issue for future works.

\subsection{Perturbed equations of motions:}
{\hskip 2em}The axial perturbations have an odd-parity $(-1)^{l+1}$ for the multi pole $l$ (this parameter is an integer) under the rotation in the two dimensional plane $(\theta,\phi)$. They can be expressed in terms of the expansion of spherical harmonics  $Y_{lm}(\theta,\phi)$, but we set $m=0$ without loss of generality. The components of odd-mode metric perturbations, in the Einstein frame, are written as
\begin{eqnarray}
ds^2=ds_{background}^2+2 Q(t,r) \sin(\theta) \partial_\theta Y_{l0}(\theta,\phi)dt d\phi +2 W(t,r)\partial_\theta Y_{l0}(\theta,\phi)\sin(\theta) dr d\phi,\label{dstot}
\end{eqnarray}
where $Q(t,r)$ and $W(t,r)$ are functions of the time $t$ and the radial coordinate $r$. Note that theirs corresponding metrics in the Jordan frame can found via the transformation: $\overline{Q}(t,r)\rightarrow \C\; Q(t,r)$ and $\overline{W}(t,r)\rightarrow \C\; W(t,r)$.  In contrast to even modes, the  perturbations of the scalar field, the energy density and the pressure are vanished in the odd modes. But the four vector velocity of the matter fluid is not zero where the only non vanishing competent is: $\delta \overline{u}^{\phi}=v(t,r)\partial_\theta Y_{l0}(\theta,\phi)/(r^2 \sin(\theta))$. In that case,  the equation of matter conservation, at the first order of perturbation, leads to
\begin{eqnarray}
v(t,r)=\frac{1}{\sqrt{\C} f}Q(t,r).\label{sv}
\end{eqnarray} 
Now, we perturb the action (\ref{S}) up to second order and after performing some integration by part, we arrive  
\begin{eqnarray}
S^{(2)}=\int dt dr \left\lbrace q_{0} \left(\dot{W}(t,r)-Q'(t,r)+\frac{2}{r}Q(t,r)\right)^2+q_{1}W(t,r)^2+q_{2}Q(t,r)^2\right\rbrace,
\end{eqnarray}
where the dot denotation represents the derivative with respect to $t$. And the coefficients $q_{1}$, $q_{2}$ and $q_{3}$ are given by 
\begin{eqnarray}
q_{0}&=&\frac{l(l+1)\cR}{4 f h},\nonumber\\
q_{1}&=&\frac{l(l+1)f\left(2 \C \C_X (3 \overline{P}-\overline{\rho} )r^2  \left(\varphi '\right)^2+h^2 \cR (l(l+1)-2)\right)}{4 h^3 r^2},\nonumber\\
q_{2}&=&\frac{l(l+1)(l(l+1)-2)\cR h}{4 r^2 f}.
\end{eqnarray}

By defining the quantity $\chi(t,r)=\dot{W}(t,r)-Q'(t,r)+\frac{2}{r}Q(t,r) $, which is a gauge invariant, and by introducing the Lagrange multiplier $\lambda(t,r)$ in the action $S^{(2)}$, we obtain
\begin{eqnarray}
S^{(2)}&=&\int dt dr \left\lbrace q_{0} \chi(t,r)^2+q_{1}W(t,r)^2+q_{2}Q(t,r)^2\right.\nonumber\\
& &\left. +\lambda(t,r)\left(\chi(t,r)-\dot{W}(t,r)+Q'(t,r)-\frac{2}{r}Q(t,r)\right)\right\rbrace .\label{S2}
\end{eqnarray}

By varying this action with respect to $\chi(t,r)$, $Q(t,r)$ and $W(t,r)$, gives, respectively, the following equations
\begin{eqnarray}
\lambda(t,r)&=&-2 q_0 \chi(t,r),\label{per1}\\
\lambda'(t,r)+\frac{2}{r}\lambda(t,r)&=&2q_2 Q(t,r),\label{per2}\\
\dot{\lambda}(t,r)&=&-2q_1 W(t,r),\label{per3}
\end{eqnarray} 

where we have assumed that $l\geqslant 2$. Since, we have $q_1=0$ outside the star for $l=1$, the function $\lambda$ will depend only on $r$ and therefore, $\chi(t,r)$  becomes a radial function. In addition, the Eq. (\ref{per2}) can be solved analytically in the vacuum as
 \begin{eqnarray}
\lambda\backsim \frac{c_1}{r^2}.
\end{eqnarray}

where $c_1$ is constant of integration. Thus, after substituting this solution in Eq.(\ref{per1}) and after fixing the gauge $W(t,r)=0$, the metric $Q(t,r)$ is written as
\begin{eqnarray}
Q(t,r)=-\frac{2 c_1}{3l(l+1)\cR r}+y(t)r^2,
\end{eqnarray}

where $y(t)$ is an arbitrary function of $t$. Providing that we have chosen the gauge $W(t,r)=0$, the function $y(t)$ must be a constant to verify the equation (\ref{per1}).\\

However, inside the star we do not need to fix the gauge because we do not have a residual gauge degree of freedom. In fact, inside the star we have only $q_2=0$ when $l=1$ which means that our metrics and Lagrange multipliers are functions of $t$ and $r$. By integrating the Eqs. (\ref{per1}), (\ref{per2}) and (\ref{per3}), we obtain
\begin{eqnarray}
\lambda(t,r)&=& \frac{y_1(t)}{r^2},\label{lambdaint}\\
Q(t,r)&=&y_2(t)r^2+r^2\int^{r}\frac{y_1(t)q_1(R)-\ddot{y}_1(t)q_0(R)}{2 R^4 q_1(R) q_0(R)}dR,\label{Qint}
\end{eqnarray}
where $y_1(t)$ and $y_2(t)$ are arbitrary functions of $t$.  In order to ensure the continuity between the outside and the inside of the star at any time, we must have $y_1(t)=c_1$ and hence the solution (\ref{Qint}) is reduced to 
\begin{eqnarray}
Q(t,r)&=&y_2(t)r^2+r^2\int^{r}\frac{c_1}{2 R^4 q_0(R)}dR.
\end{eqnarray}

We notice that this solution is written independently from the matter and the function $A$. In the next sections, we will consider only the case   $l\geqslant 2$ in which the conformal coupling appears in the equations of motion.
\subsection{Ghost and Laplacian instability:}
{\hskip 2em}In order to derive the conditions of Ghost and Laplacian instability, we solve the equations (\ref{per1}), (\ref{per2}) and (\ref{per3}) for $\chi$, $Q$ and $W$, then we insert the result in the action (\ref{S2}). After integrating by part, we get

\begin{eqnarray}
S^{(2)}&=&-\int dt dr\frac{1}{4} \left\lbrace \frac{1}{q_1} \dot{\lambda}(t,r)^2+\frac{1}{q_2} \lambda'(t,r)^2+\left(\frac{2 \left(r q_2'+3 q_2\right)}{q_2^2 r^2}+\frac{1}{q_0}\right)\lambda(t,r)^2\right\rbrace. \label{S3}
\end{eqnarray}
\\
For $l\geqslant 2$, the ghosts are absent when $q_2$ is positive. At the exterior of the star this condition is satisfied, since $f$ and $h$ are positive, but inside the neutron star $q_2>0$ is satisfied when
\begin{eqnarray}
2 \C \C_X (3 \overline{P}-\overline{\rho} )r^2  \left(\varphi '\right)^2+h^2 \cR (l(l+1)-2)>0.\label{ghost}
\end{eqnarray}

\begin{figure}[h]
\centering
\includegraphics[scale=0.8]{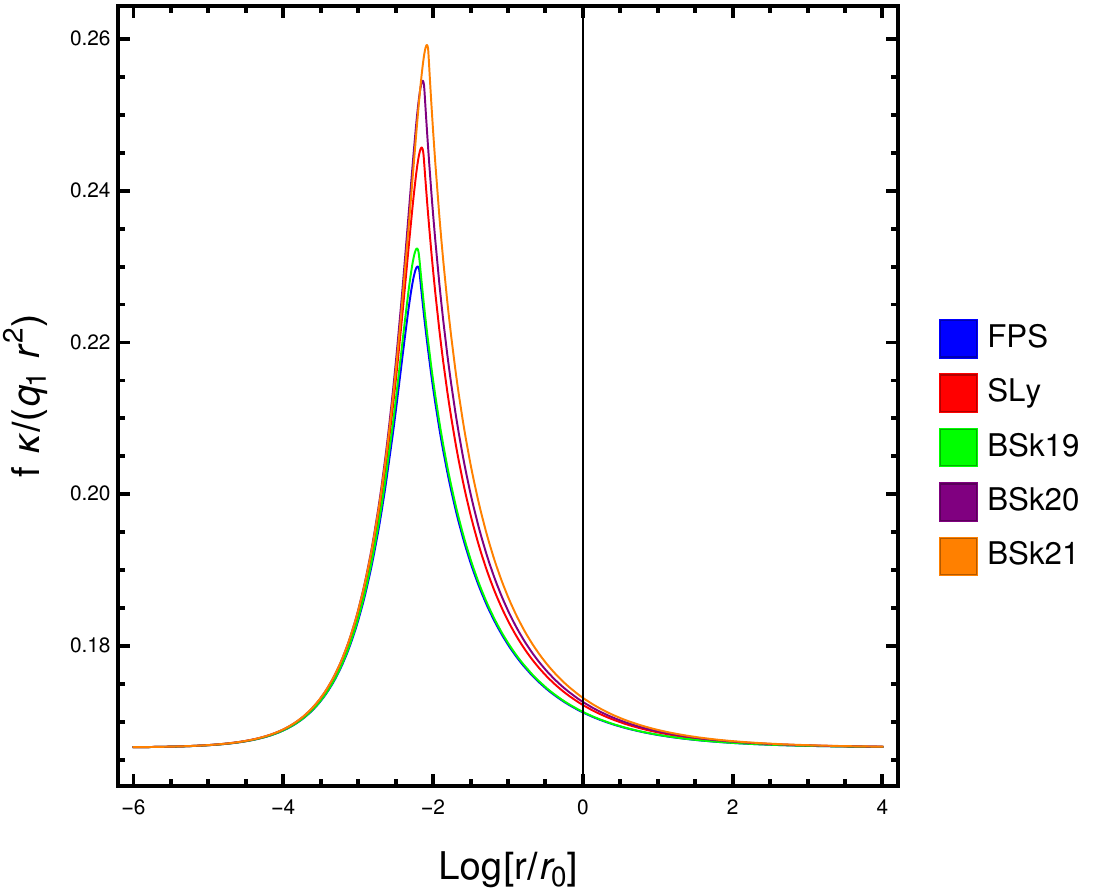} 
\caption{\small We show the ghost instability of the neutron star using different EOSs, for the case $\overline{\rho}(r=0)=10\rho_0$ and $\tilde{\cC}=0.03$.}
\label{C1}
\end{figure}

This condition is also satisfied at the center of the star because we find that $4 q_2 r^2\simeq l(l+1)(l(l+1)-2)\cR$ when $r\ll r_s$. In addition, in the case of (\ref{function}) the model does not have the ghosts instability wherever in the space, as it has been shown in the Fig.\ref{C1}. We point out that the ghost instability, for the axial modes, is absent in the GR case and this can been seen directly from (\ref{ghost}) by setting $\C=1$.

\begin{figure}[h]
\centering
\includegraphics[scale=0.8]{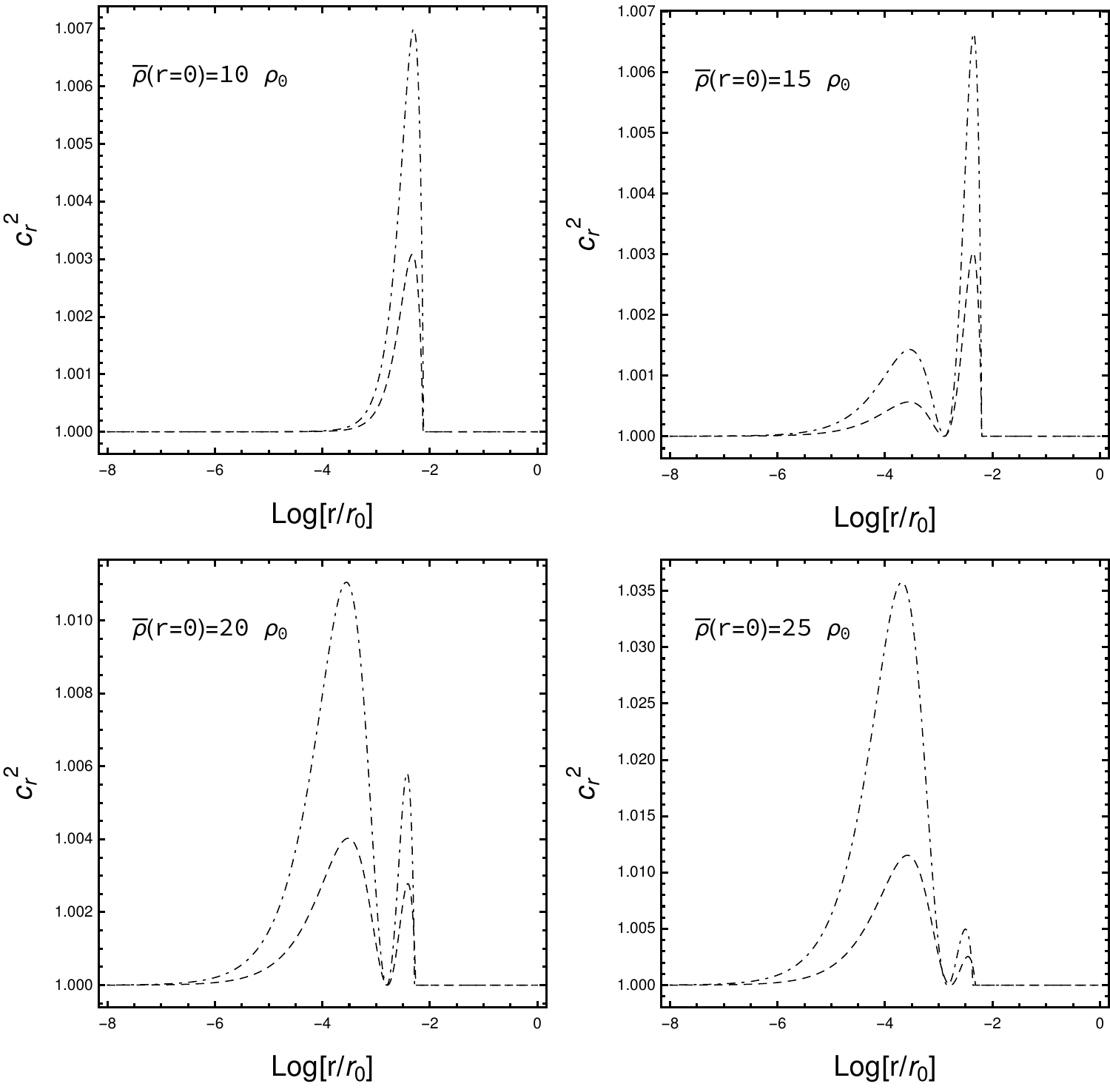} 
\caption{\small We show the Laplacian instability of the neutron star using the realistic EOS SLy, for the cases $\tilde{\cC}=\{0.03\; (\text{Dashed curve}),\;0.045\; (\text{DotDashed curve)}\}$.}
\label{cr}
\end{figure}

The conditions Laplacian instabilities are obtained by inserting the solution of the form 
\begin{eqnarray}
\lambda(t,r)=e^{-I (\omega t- k r)},
\end{eqnarray}
in the action (\ref{S3}), where $\omega$ and $k$ are the frequency and the wave number, respectively. By taking the limits $\omega\rightarrow\infty$ and $k\rightarrow\infty$, the Laplacian instabilities are absent when $\tilde{c}_r^2=(k/\omega)^2=-q_2/q_1$ is positive. In fact, this instability is satisfied if the condition $q_2>0$, which is the case in our model. We can use the particular model used before to simulate the behaviour of the speed propagation $c_r^2=\tilde{c}_r^2 h^2/f^2$ of $\lambda(t,r)$ in proper time as it has been plotted in the Fig.\ref{cr}. The numerical results do not only  confirm that our model is free from Laplacian instability but they also show that propagation speed in the radial direction is equivalent to the speed of light at the center and at infinity. We notice that $c_r$ deviates from $c$ at the surface of the star with $3.5\%-0.07\%$ where the deviation is more important by  increasing the central density. 

Finally, the propagation speed square in the  angular direction $c_\Omega^2$ is derived by using the solution $e^{-I (\omega t- l\theta)}$ and taking the limits $l\rightarrow\infty$ and  $\omega\rightarrow\infty$. Following the same steps of the calculation of $c_r^2$, we arrive to $c_\Omega^2=1$ in proper time and hence the Laplacian instability is absent along the angular direction.

\section{Quasi-Normal modes and universal relations:}\label{Sec3}
{\hskip 2em}In this section, we wish to present our numerical  results by simulating the perturbed equations of motion. We will see the behaviour of  quasi-normal modes by varying different physical characteristic of the neutron star and by varying also the parameters in the model. At the end of this section, the universal relations between the compactness $C=M/r_s$ and the quasi normal are derived.
\subsection{Asymptotic behaviour}
{\hskip 2em}The equations of motion are calculated by substituting Eq.(\ref{per1}) in Eqs.(\ref{per2}) and (\ref{per3}). The function $\chi$ is eliminated by using its definition. After some calculations, we arrive to the following equations, in Fourier space ($F(t,r)=\int d\omega e^{-I \omega t} F(r)$) and here $\omega=\omega_R+I \omega_I$ is a complex impulsion,
 \begin{eqnarray}
Q'&=&\frac{i q_2 W \omega }{q_1}-\frac{Q \left(r q_1'+2 q_1\right)}{q_1 r},\label{per12}\\
W'&=&\frac{2 W}{r}-\frac{i Q \left(q_0 r \omega ^2+q_1 r\right)}{q_0 r \omega }.\label{per22}
\end{eqnarray}

Here, the functions $W$ and $Q$ depend only on $r$. One can also combine these equations to get a differential  equation  of second order and it is calculated by
\begin{eqnarray}
W''+W' \left(\frac{q_1'}{q_1}+\frac{2}{r}\right)+W \left(\frac{q_1''-q_2 \omega ^2}{q_1}-\frac{q_1'^2}{q_1^2}-\frac{q_2}{q_0}-\frac{2}{r^2}\right)=0.\label{peroerder2}
\end{eqnarray}

Now, we will use the form of the functions (\ref{function}) to study the asymptotic behaviours of the perturbed metrics. Perturbing the metrics $Q$ and $W$ around the center of star as, respectively, $Q\approx Q_0 r^{l+1}$ and $W\approx W_0 r^{l+1}$ as and inserting them in Eqs.(\ref{per12}) and (\ref{per22}), we found that $W$  behaves asymptotically as 
\begin{eqnarray}
W&=&-\frac{I Q_0 \omega }{l+2}r^{l+2},\label{Wper}
\end{eqnarray}
where the parameter of the model $\cC$ is absent. Thus, our model, at the perturbed level, is identical to GR when we are close to the center of the star. At large value of $r$, the equations of perturbation corresponding to our model are also identical to GR and the analytical solution when $r$ tends to infinity is written as
\begin{eqnarray}
W\approx r \left(A_{out} e^{I \omega r^{*}}+A_{in} e^{-I \omega r^{*}}\right),
\end{eqnarray}
where $A_{out}$ and $A_{in}$ are complex constants, and $r^{*}$ is  the tortoise coordinate defined  by $dr^{*}= (h/f)dr$. This solution is composed into two part where the first part is the outgoing wave and the second one is the ingoing wave. If we impose  $A_{in}=0$, we will have a purely outgoing wave at infinity which is possible for discrete values of $\omega$.

\subsection{Numerical method}
{\hskip 2em}The numerical solutions are obtained by integrating Eqs. (\ref{per12}) and (\ref{per22}) from the center of the star to large value of $r$. To have physical solutions, the perturbed metric must be regular at the center, must be continued at the surface and must be an outgoing solutions at infinity. The two first conditions can be satisfied by considering the asymptotic behaviours  (\ref{Wper}) as initial conditions and by considering $Q_{ext}(r=r_s)=Q_{int}(r=r_s)$ and $W_{ext}(r=r_s)=W_{int}(r=r_s)$. However, because of the exponential divergence that appears in the ingoing wave, the third condition can not be satisfied numerically. In order to overcome this problem, we will use the complex coordinate $y$, defined by
\begin{eqnarray}
r=r_s+ y e^{I \Phi},
\end{eqnarray}
where $\Phi$ is a constant which is chosen to satisfy the condition $\omega_R \sin(\Phi)+ \omega_I \cos(\Phi)<0$. Indeed, this later will enforce the solution to be a purely outgoing wave and thus numerical divergence problems are avoided. \\

\begin{figure}[h]
\centering
\includegraphics[scale=0.7]{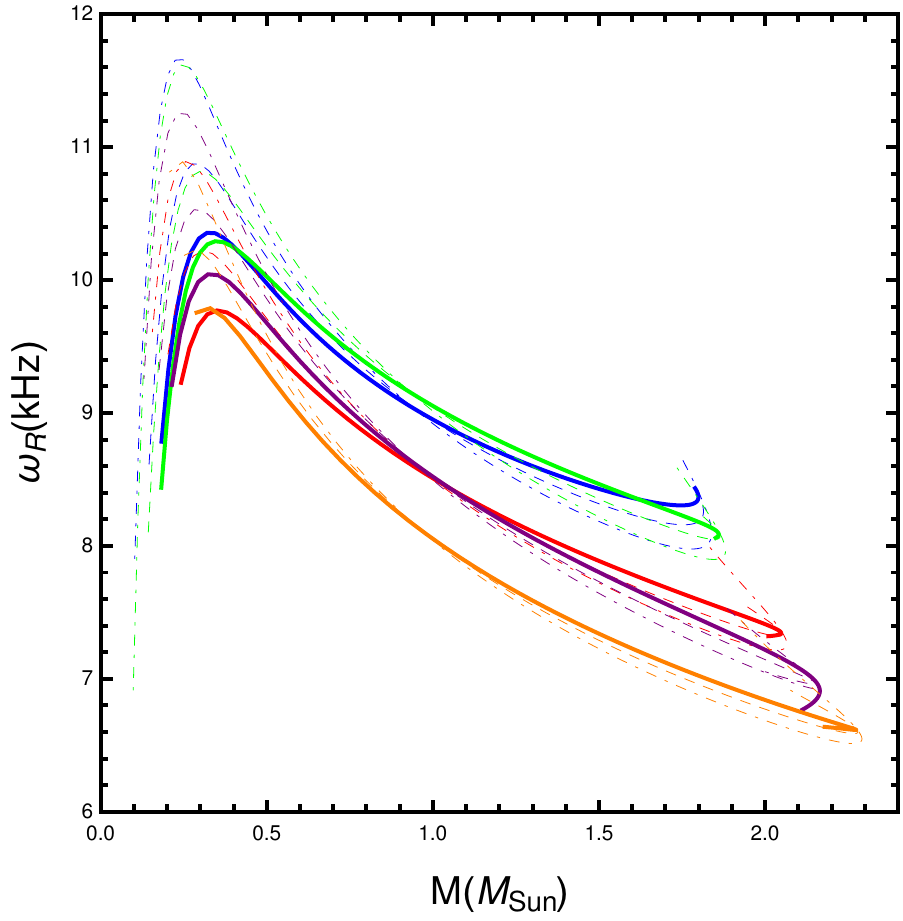} 
\includegraphics[scale=0.7]{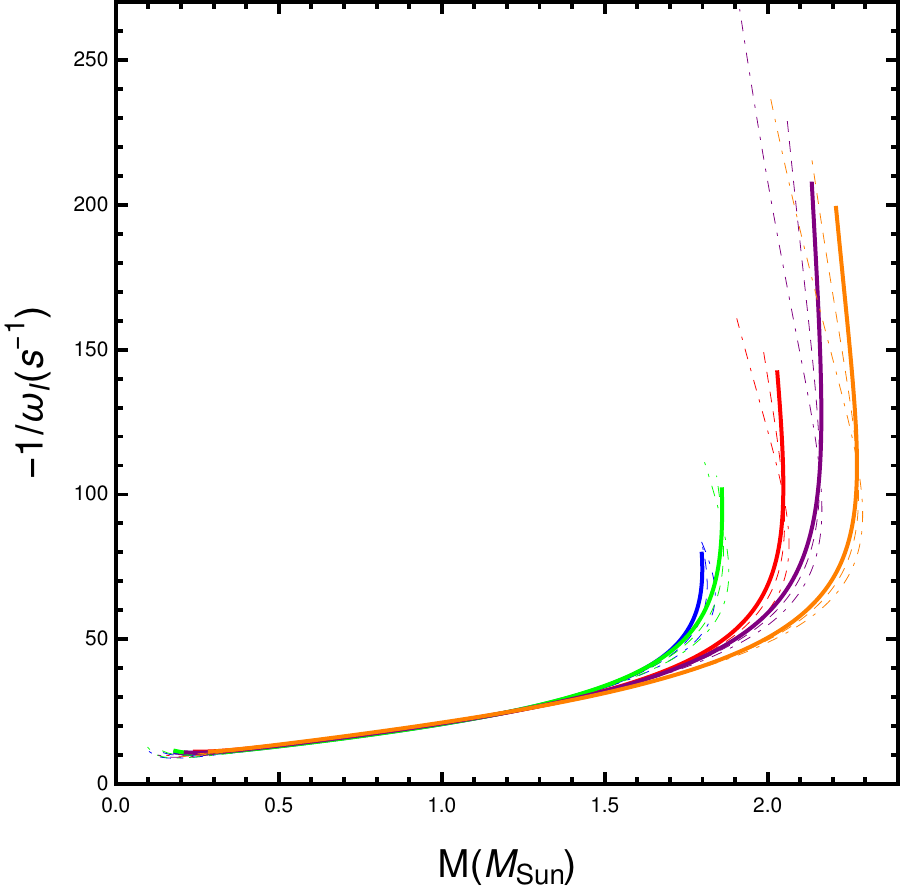} 
\caption{\small \textit{left graph:} The variation of the frequency $\omega_R$ with respect to the mass of the neutron star. \textit{Right graph:} We plot the damping time $-1/\omega_I$ as a function of $M$. In the both graphs, we used the same color and curves for the same cases plotted in the Fig.\ref{MRs}.}
\label{wM}
\end{figure}

The numerical integration of the Eqs. (\ref{per12}) and (\ref{per22}) is performed on one hand from the center to the surface along the real axes, by taking into consideration the regularity at the center, and on the other hand from $R_{out}=80 r_s$ to $r_s$ along the axes with slope $\arctan\left[-\omega_I/\omega_R\right]$, by using the series development
\begin{eqnarray}
W(r)=\left(1-\frac{2M}{r}\right)^{-1} r e^{I \omega r^{*}}\sum_{n=0}^{n=15}c_n^{*}r^{-n},
\end{eqnarray}
as initial condition at $R_{out}=80r_s$. Note that $c_n^{*}$ are constants which can be determined from Eq.(\ref{peroerder2}). The continuity at the surface is expressed as
\begin{eqnarray}
\frac{W'(r_s)}{W(r_s)}=\frac{1}{e^{I \Phi}}\frac{W'(y=0)}{W(y=0)},
\end{eqnarray} 

where this equation is true only for discrete values of $\omega$.  In Fig.\ref{wM}, we plot the frequency $\omega_R$ and the damping time  $\tau=-1/\omega_I$ as a function of the mass, for $l=2$ and $\tilde{\cC}=\{0,\, 0.03,\,0.045\}$ by using different kind of realistic EOSs. We observe that depending on the EOS and the parameter $\cC$, we get different variations of $\omega_R$  and $\tau$ with respect to $M$. In other words, the quasi-normal modes  are affected by the matter present inside the neutron star and by the extra degree of freedom (scalar field) proposed in our model. However, the damping time is not affected by the parameter $\cC$ and the kind of EOS when the mass is $ M <1.4 M_{Sun}$ then the distinction among each case is being observable when $ M >1.5 M_{Sun}$, as it has been shown in the the right graph.\ref{wM}. In addition, we notice that the deviation of the damping time in our model is important   when the mass of the neutron star is close to the minimum radius. Finally, the maximums of $\omega_R$  and $\tau$  in our model, for each EOS, are larger that the ones in GR where  they increase by decreasing the value of $\cC$.

\subsection{Universal relations}\label{sec33}
{\hskip 2em}
We close this section by showing the universal relations, discussed in Refs.\cite{AltahaMotahar:2019ekm,Blazquez-Salcedo:2018qyy,Blazquez-Salcedo:2018tyn,Blazquez-Salcedo:2013jka,Blazquez-Salcedo:2012hdg}, for the  frequencies $\omega_R$ and the damping times $\omega_I$. These universal relations relate the QNMs to the physical features of the neutron star such as: mass, compactness or radius. In addition, they can be used to show the effects of the conformal coupling and the kind of EOS on their behaviours. The relations are derived by fitting the data obtained from the numerical solutions of the our model and GR for different type of EOSs.  \\

\begin{figure}[h]
\centering
\includegraphics[scale=0.6]{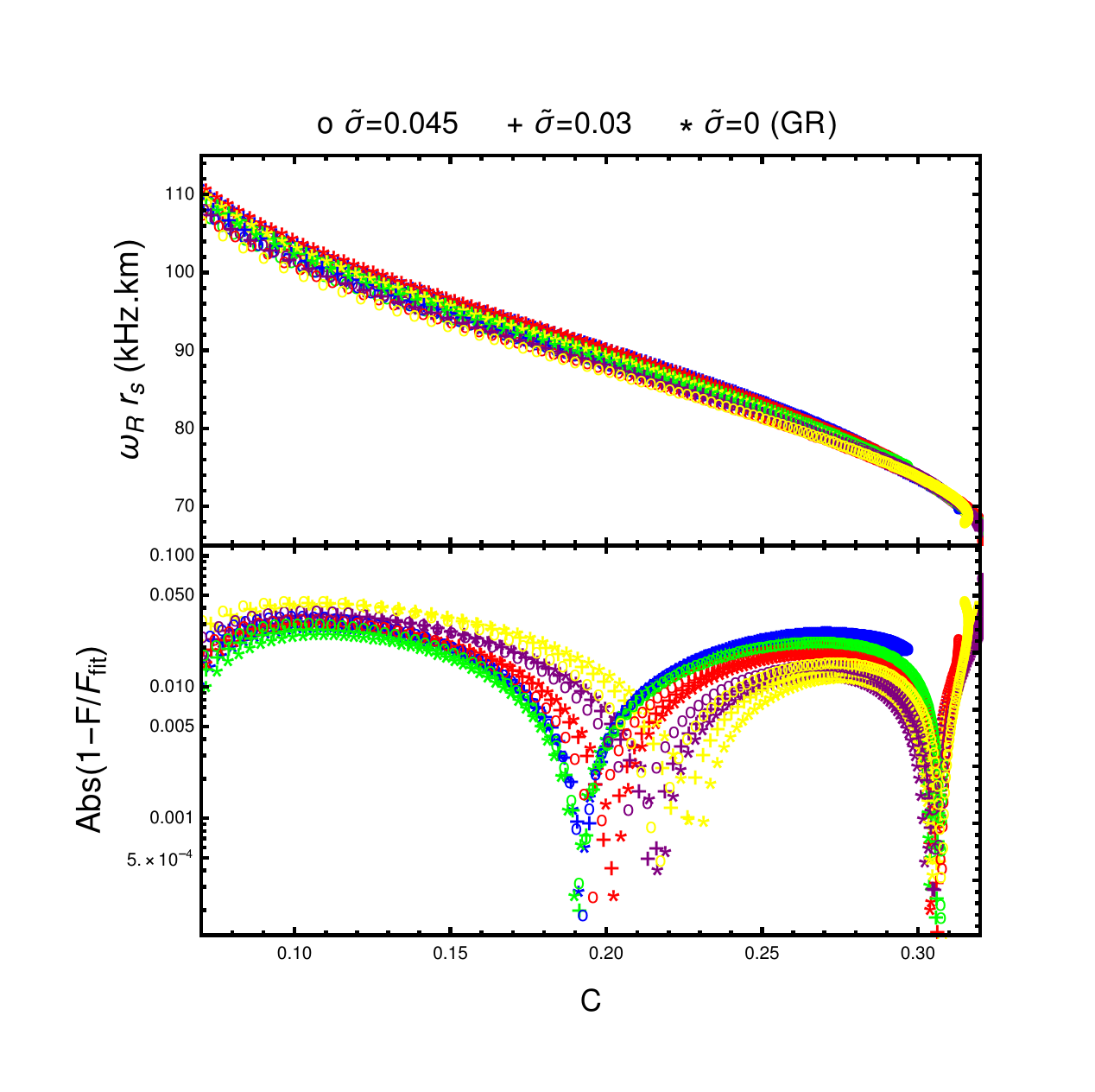} 
\includegraphics[scale=0.6]{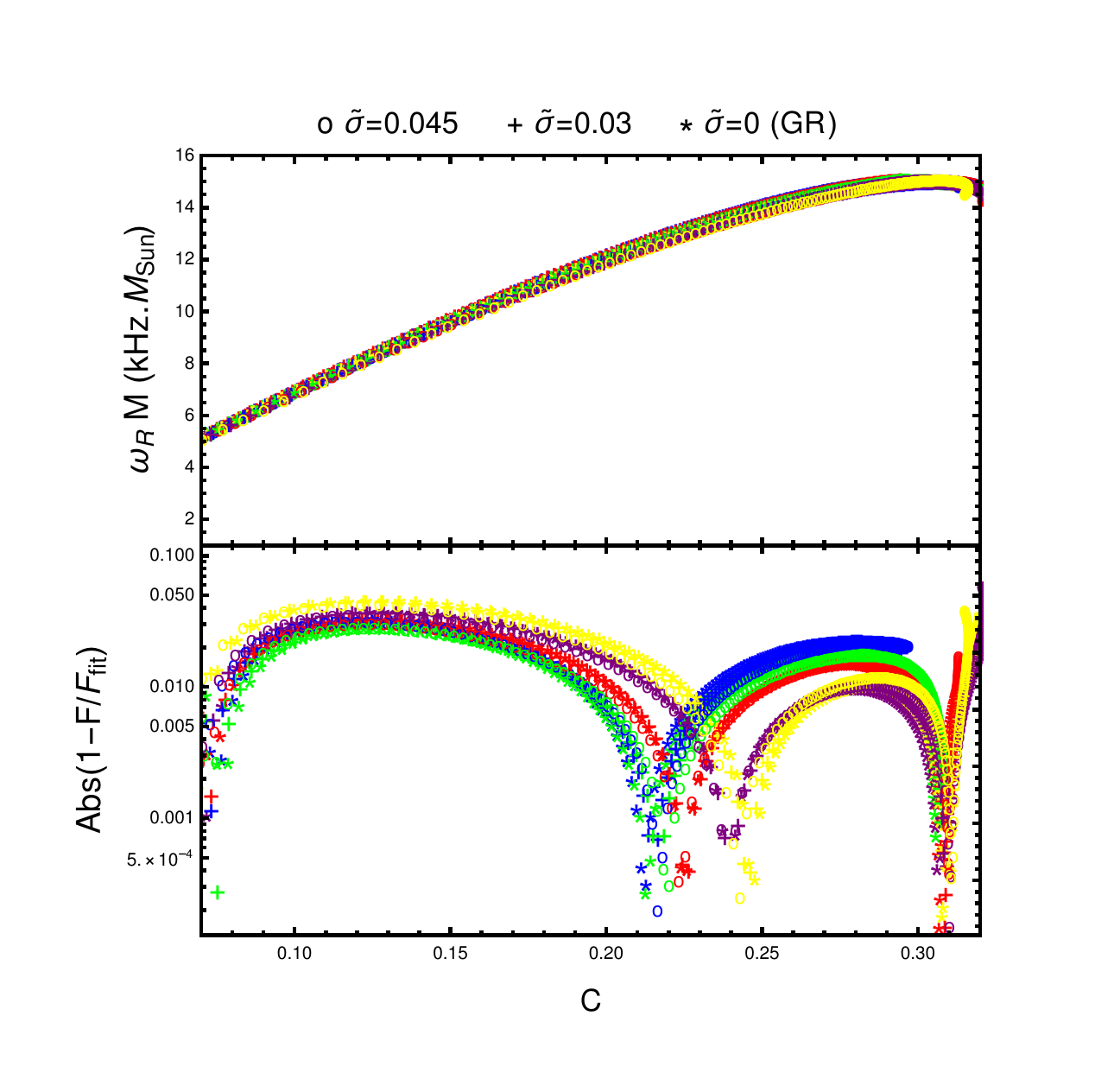} 
\includegraphics[scale=0.6]{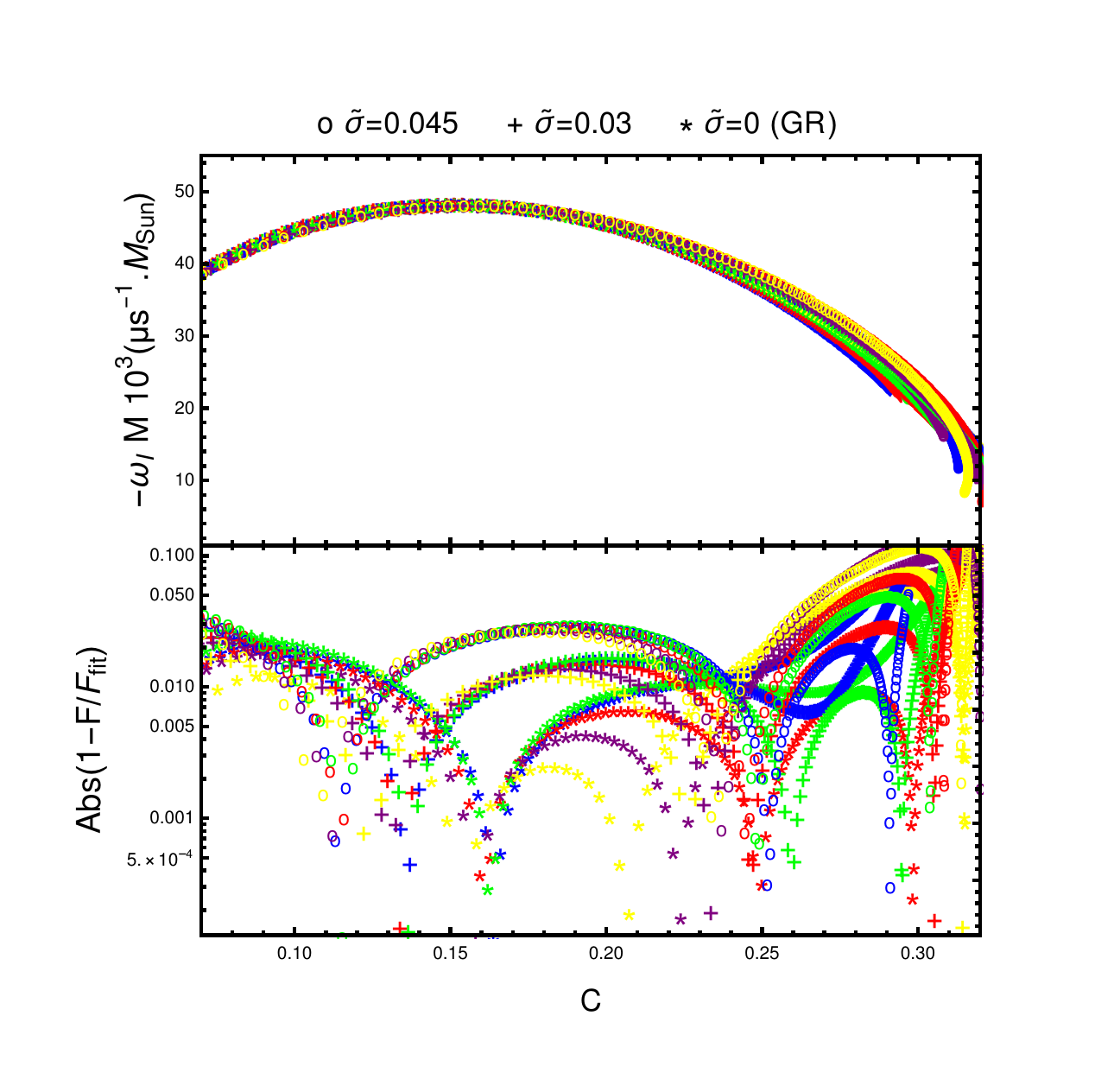}
\includegraphics[scale=0.6]{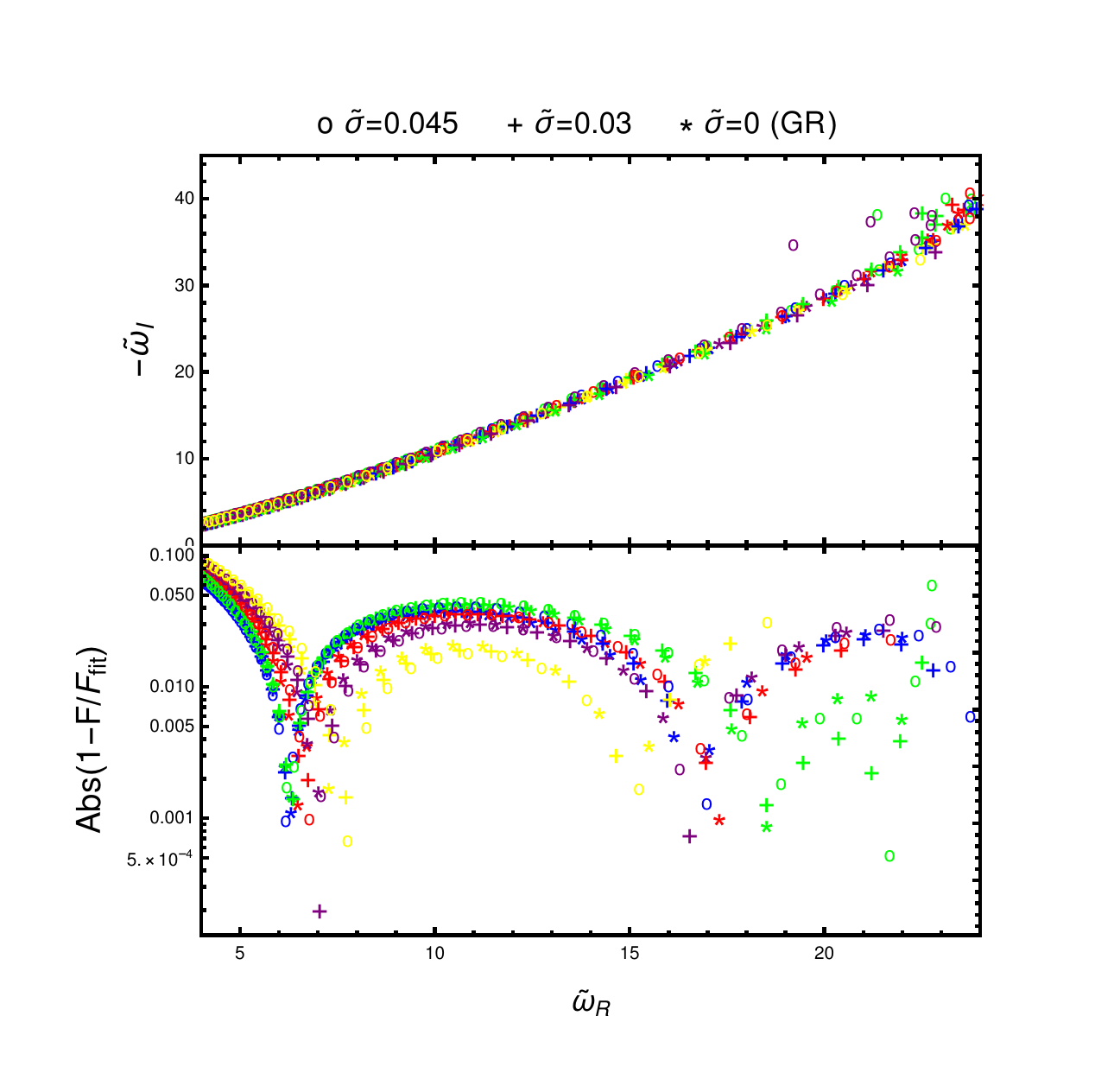}
\caption{\small  We plot the universal relations discussed in the subsection. \ref{sec33} for different value of $\tilde{\cC}$ and for the EOSs: FPS (Blue), SLy (Red), BSk19 (Green), BSk20 (Purple), BSk21 (Yellow).}
\label{wrC}
\end{figure}

 We consider the  universal relation in which the normalized frequencies $r_s\omega_R$ are quadratic function of the compactness of the star. Our numerical fitting with the model is summarized as
 \begin{eqnarray}
 \omega_R . r_s =\left\{  
    \begin{array}{ccc}
     124.587 - 176.553 \;C + 22.592 \;C^2 & \quad\tilde{\cC}=& 0\,(GR), \\
     124.464 - 187.851 \;C + 59.600 \;C^2 & \quad\tilde{\cC}=& 0.03,  \\
     123.631 - 197.827 \;C + 100.52 \;C^2 & \quad\tilde{\cC}=& 0.045, 
    \end{array}\right.
 \end{eqnarray}
were $C=M/r_s$ is compactness of the neutron star. We can also use the normalized  frequencies $M\omega_R$ to fit  our model with the data for the same function form. In this case, the universal relation is given by 
\begin{eqnarray}
 \omega_R . M =\left\{  
    \begin{array}{ccc}
     -0.375 + 88.758 \;C - 125.704 \;C^2 & \quad\tilde{\cC}=& 0\,(GR), \\
     -0.286 + 86.532 \;C - 119.627 \;C^2 & \quad\tilde{\cC}=& 0.03,  \\
     -0.172 + 83.359 \;C - 110.594 \;C^2 & \quad\tilde{\cC}=& 0.045. 
    \end{array}\right.
 \end{eqnarray}
We observe, in the two cases, the deviations from GR by varying  the constants $\tilde{\cC}$ which is expected from our analyses in the last sections. Moreover, we can see that our model is well fitted with the universal relation, since the deviation is less than $10\%$ as it have been shown in the top of the figure \ref{wrC}. In the bottom of this figure, we plot the  relations damping time-compactness and $\tilde{\omega}_I$-$\tilde{\omega}_R$, where $\tilde{\omega}=\omega/\sqrt{\overline{P}_c}$, for the five EOSs used in the second section. By fitting our model with the numerical results, the universal relation $M/\tau$-$C$ of the quadratic form reads
\begin{eqnarray}
 \frac{M}{\tau}.10^3=\left\{  
    \begin{array}{ccc}
     16.893 + 403.849 \;C - 1294.31 \;C^2 & \quad\tilde{\cC}=& 0\,(GR), \\
     14.770 + 429.469 \;C - 1356.09 \;C^2 & \quad\tilde{\cC}=& 0.03,  \\
     12.007 + 464.757 \;C - 1445.61 \;C^2 & \quad\tilde{\cC}=& 0.045. 
    \end{array}\right.
 \end{eqnarray}
And the phenomenological relation $\tilde{\omega}_I$-$\tilde{\omega}_R$ is calculated as
\begin{eqnarray}
 \frac{M}{\tau}.10^3=\left\{  
    \begin{array}{ccc}
     -0.839+ 0.761 \;C + 0.0372\;C^2  & \quad\tilde{\cC}=& 0\,(GR), \\
     -0.870+ 0.779 \;C + 0.0370 \;C^2 & \quad\tilde{\cC}=& 0.03,  \\
     -0.910+ 0.978 \;C + 0.0362 \;C^2 & \quad\tilde{\cC}=& 0.045. 
    \end{array}\right.
 \end{eqnarray}
These relation have deviations less than $10\%$ when $0.1<C<0.25$ which can be observe in Fig.\ref{wrC}. Therefore, one can say that our model verifies the universal relations, which may let's us think that probing the conformal coupling constant will be possible with future observations of gravitational waves. Furthermore, the deviation of QNMs of the model from those in GR is more important at high density than at low density, and thus observing neutrons star with high density will allow us to have a better constraint on $\tilde{\cC}$.

\section{Conclusion:}\label{Sec5}
{\hskip 2em}In this study, we have proposed to compute the axial QNMs of neutron stars for five realistic EOSs where the matter is couple to metric with a conformal transformation. We have used the spatial case in which the conformal function has the form (\ref{function}). To do so, we have computed the perturbed equations that correspond to the metric (\ref{dstot}) then we have performed a numerical integration of the resulting equations. The conditions of avoiding ghost and Laplacian instability for the axial perturbation have been also calculated.

{\hskip 2em}Indeed, our model is free from these instabilities in every point of the space time. At the center of the star and at infinity, the  propagation speed  of the gravitational waves are equal to the speed of light and it is not affected by the parameter $\cC$. Furthermore, our numerical analysis showed  that the constant $\cC$ affects the behaviour of mass, radius and the axial quasi-normal modes. The deviation of the frequency and the damping time in our model from those in GR increases by increasing $\cC$. However, the damping time with $\tilde{\cC}=\{0\;,0.03\;,\;0.045\; \}$ are quasi identical at low masses, for the five EOSs, but it is not the case for large masses. These results have motivated us to extend our work and confront the simulation with the universal relations of Refs.\cite{AltahaMotahar:2019ekm,Blazquez-Salcedo:2018qyy,Blazquez-Salcedo:2018tyn,Blazquez-Salcedo:2013jka,Blazquez-Salcedo:2012hdg}. Ones again, the constant $\cC$ effects the coefficients of the quadratic universal relations. We have shown that for the values $\tilde{\cC}=\{0\;,0.03\;,\;0.045\; \}$ and for the EOSs that we have used in this paper, the universal relations deviations are less than $10\%$. Therefore, it would be interesting to consider the polar perturbation but we leave this for future works.

\bibliographystyle{ieeetr}
\bibliography{bibliography}

\end{document}